\documentstyle[12pt]{article}
\begin{document}
\begin{flushright}
 UB--ECM--PF 96/18\\
 ULB--TH--96/17\\
 hep-th/9611056\\
\end{flushright}
\vspace{2ex}

\begin{center}
 {\large{\bf Ward identities for rigid symmetries of higher order}}
\end{center}
\vspace{2ex}

\begin{center}
 Friedemann Brandt\,$^{a}$,
 Marc Henneaux\,$^{b,c}$\ and
 Andr\'{e} Wilch\,$^{b}$
\end{center}
\vspace{2ex}

\begin{center}{\sl
 $^a$ Departament d'Estructura i Constituents de la Mat\`eria,
 Facultat de F\'{\i}sica,
 Universitat de Barcelona,
 Diagonal 647, E--08028 Barcelona, Spain.\\[1.5ex]

 $^b$ Facult\'e des Sciences, Universit\'e Libre de Bruxelles,\\
 Campus Plaine C.P. 231, B--1050 Bruxelles, Belgium.\\[1.5ex]

 $^c$ Centro de Estudios Cient\'\i ficos de Santiago,\\
 Casilla 16443, Santiago 9, Chile.
}\end{center}
\vspace{2ex}

\begin{abstract}
Higher order conservation laws, associated with conserved 
antisymmetric tensors $j^{\mu_1 \ldots \mu_k}$ fulfilling
$\partial_{\mu_1} j^{\mu_1 \ldots \mu_k} \approx 0$,
are shown to define rigid symmetries of
the master equation.  They thus lead to independent Ward identities
which are explicitly derived.  
\end{abstract}
\vspace{3ex}

\noindent
Gauge theories may possess global (= rigid) symmetries
in addition to their local (= gauge) symmetries. Through Noether's first
theorem, non-trivial rigid symmetries of the classical action correspond
to non-trivial conserved currents,
\begin{equation}
\partial_\mu j^\mu \approx 0\;.
\label{cons1}
\end{equation}
Now, (\ref{cons1}) is actually only a special case of the more general 
conservation law
\begin{equation}
\partial_{\mu_1} j^{\mu_1  \dots \mu_k}  \approx 0
\label{cons2}
\end{equation}
where  $j^{\mu_1  \dots \mu_k}$ is  completely
antisymmetric and the symbol
$\approx$ denotes weak (i.e. on-shell) equality.  Non-trivial
solutions of (\ref{cons2}) define what we
call non-trivial conservation laws of order $k$ \cite{comment0}.
Although it has been proved under fairly general conditions
that non-trivial conservation laws of higher order $k>1$
are absent for theories without gauge invariance
\cite{Vinogradov,Bryant,bbh1},
they may be present in the case of gauge theories.
Examples are given by $p$-form gauge theories which admit
non-trivial conserved antisymmetric tensors of rank $p+1$ \cite{hks}.
These conservation laws play an important role in supergravity 
\cite{Page}.  

The quantum mechanical implications of the rigid symmetries that are 
associated with ordinary conserved currents (\ref{cons1}) 
are well understood.  If these symmetries are non-linear, they get 
renormalized.   The most expedient way to derive the corresponding Ward
identities is to introduce sources for the composite operators 
representing the variations of the fields \cite{ZJ}.  In order to avoid an
infinite number of such sources (one for the first
variation, one for the second variation etc. \cite{Bonneau}),
constant ghosts are introduced of ghost number $1$ and of 
Grassmann parity opposite to that of the symmetry parameter \cite{Blasi}. 
The Ward identities then follow by solving an extended master equation
\cite{Blasi,Piguet1,Piguet2},
the explicit form of which will be given below.  This approach has
proved useful in the investigation of the renormalization
and anomaly problems in globally supersymmetric models \cite{many}.
However, it has been shown that the construction of
a local solution of the extended master equation may get obstructed --
already at the classical level --
whenever there exist higher order non-trivial conservation laws \cite{global}.
When this occurs, it may not be possible to incorporate the standard rigid
symmetries along the lines of \cite{Blasi}.

The purpose of this letter
is to show that the obstructions can be avoided if one extends
the approach of \cite{Blasi} by properly including in the formalism also
the higher order conservation laws.  We  then derive the
Ward identities associated with these
higher order conservation laws and show that the analysis of the
anomalies in both the rigid and gauge symmetries may still be 
formulated as a cohomological problem. 

The crucial observation that enables one to include 
the higher order conservation laws is that these latter
define  {\it additional rigid symmetries of the theory}. 
This fact does not contradict Noether's
theorem, which associates
the rigid symmetries of the classical action with ordinary conserved currents.
Rather, it extends this theorem because the additional rigid symmetries are
symmetries of the solution of the master equation.

Our starting point is thus the solution $S=S[\Phi^a,\Phi^*_a]$
of the master equation for
the gauge symmetries \cite{bv},
\begin{equation}
(S,S) = 0,
\label{master1}
\end{equation}
where $(\ ,\ )$ is the standard antibracket.
The $\{\Phi^a\}$ are the fields
(classical fields, ghosts for the
gauge symmetries, ghosts of ghosts if
necessary, antighosts, Nakanishi-Lautrup auxiliary fields) while the
$\{\Phi^*_a\}$ are the corresponding antifields. The
master equation (\ref{master1}) always admits a solution
$S$ which is a {\em local} functional \cite{hen91}.

As shown in \cite{bbh1}, one can associate with each 
conservation law 
$\partial_{\mu_1} j_A^{\mu_1  \dots \mu_k}  \approx 0$ 
of order $k$ a local functional $S_A[\Phi^a, \Phi_a^*]$ 
which (i) has ghost number $-k$; and
(ii) is BRST-invariant.  Since the BRST transformation $s$ is generated
in the antibracket by $S$, (ii) means 
\begin{equation}
s\,  S_A \equiv (S_A,S) =0.
\label{inv2}
\end{equation}
But this condition expresses at the same time 
that the solution $S$ of the master equation
is invariant under the canonical transformation generated in
field-antifield space by $S_A$,
\begin{equation}
\delta_A S  \equiv (S,S_A) =0.
\label{inv1}
\end{equation}
Consequently, each conservation law defines indeed 
a symmetry of $S$.

The relationship
between the conserved antisymmetric tensors and the generators 
$S_A$  has been given in \cite{bbh1}:
since $S_A=\int d^n x \, m_A$ is BRST invariant, its
integrand satisfies
\begin{equation}
sm_A + \partial_\mu m_A^\mu = 0.
\label{cons3}
\end{equation}
If $S_A$ has ghost number $-1$ (corresponding to ordinary rigid symmetries),
$m_A^\mu$ has ghost number zero and the antifield independent part of
(\ref{cons3}) reproduces (\ref{cons1}) because the antifield
independent part of $sm_A$ vanishes on-shell.
For $k=2$, the relation (\ref{cons3}) also holds,
but now, $m_A$ has ghost number $-2$ and
$m_A^\mu$  has ghost number $-1$.  To get the conservation law
in the form (\ref{cons2}), the descent equation technique has to be used,
i.e. the differential $s$ has to be applied to (\ref{cons3}).  
Following standard arguments, this yields 
\begin{equation}
sm_A^\mu + \partial_\nu m_A^{\mu \nu} =0
\label{cons4}
\end{equation}
for some antisymmetric tensor $m_A^{\mu \nu}$ of ghost number
zero.  The conservation law (\ref{cons2}) is just the
antifield-independent part of (\ref{cons4}), since
the antifield-independent part of $sm_A^\mu$ vanishes on-shell.
Similar arguments hold for the subsequent conservation laws with $k>2$.
By using the
antifield-BRST formalism, one can consequently provide a unified treatment
for all conservation laws.
(How to deal with descent equations
in the quantum theory is discussed in \cite{Lucchesi,Piguet1}).

The symmetry generators $S_A$ possess an interesting algebraic
structure.  Assume that $\{S_A\}$ is a basis
of symmetry generators, in the sense that (i) the most general
local functional with negative ghost number generating
a symmetry of $S$
is a linear combination of the $S_A$ with constant coefficients
$\lambda^A$ up to a BRST-exact term,
\begin{equation}
(S,M)=0, \, gh(M)<0 \; \Leftrightarrow \;M = \lambda^A S_A + (S,K);
\label{basis1}
\end{equation}
and (ii) the $S_A$ are independent in cohomology,
\begin{equation}
\lambda^A S_A + (S,K) = 0 \; \Leftrightarrow \; \lambda^A = 0.
\label{basis2}\end{equation}
Since the antibracket of two $S_A$ is BRST-closed,
it must be of the form
\begin{equation}
(-1)^{\varepsilon_A}(S_A,S_B)=f_{AB}^D S_D+(S,S_{AB}) 
\label{16a}\end{equation}
for some constants $f_{AB}^C$ and some local functionals
$S_{AB}$ ($\varepsilon_A+1$ denotes the Grassmann parity of
$S_A$).  Taking the antibracket of this expression with $S_C$
and using the Jacobi identity for the antibracket then leads to
\begin{equation}
S_Ef_{D[A}^Ef_{BC]}^D=
\left(S,(-)^{\varepsilon_B}(S_{[B},S_{CA]})-S_{D[A}f_{BC]}^D\right)
\label{basis4}
\end{equation}
where $[\;]$ denotes graded antisymmetrization.
According to (\ref{basis2}), both sides of (\ref{basis4}) have to 
vanish separately.
This yields the Jacobi identity for the structure constants $f^C_{AB}$ 
and -- due to (\ref{basis1}) --
the additional identity
\[
(-)^{\varepsilon_A}(S_{[A},S_{BC]})=S_{D[C}f_{AB]}^D+
\mbox{\small{$\frac{1}{3}$}}f_{ABC}^DS_D
+\mbox{\small{$\frac{1}{3}$}}(S,S_{ABC})
\]
for some second order structure constants $f_{ABC}^D$ and
some local functionals $S_{ABC}$.
If there does not exist any higher order conservation law (and thus no
$S_A$ with $gh(S_A)<-1$), then higher order structure constants like
$f_{ABC}^D$ cannot occur.
This follows from a mere ghost number counting argument.  
However, in the presence of higher order symmetries, terms of the
form $f_{ABC}^DS_D$ are allowed and indeed do occur in explicit
examples \cite{global}.
It is clear that
the algebra of the $S_{A_1 \dots A_r}$, established
through their (antisymmetrized) antibrackets, can involve
third and higher order structure constants satisfying
generalized (higher order) Jacobi identities.
We shall not be more explicit
about this algebra here, since it is automatically incorporated
(to all orders) in the extended master equation given below.

A subset of symmetry generators $S_\alpha$ defines a
subalgebra if and only if the relations to which they lead never involve
the other symmetry generators $S_\Delta$, $A=(\alpha, \Delta)$. 
An equivalent condition is that
the structure constants $f_{\alpha_1 \alpha_2}^\Delta$,
$f_{\alpha_1 \alpha_2 \alpha_3}^\Delta$ $\dots$ all vanish.
The subset
$\{S_\alpha$, $S_{\alpha_1 \alpha_2}$, $S_{\alpha_1 \alpha_2 \alpha_3}$,
$\dots\}$ is then a closed set for the generating equations
(\ref{16a}) and the subsequent ones.  As
shown in \cite{global}, the set of all symmetry generators of
order one (standard rigid symmetries) may not form a subalgebra in
the above sense.

The fact that all the conservation laws, including the
higher-order ones, appear as symmetries of the solution
of the master equation, makes it possible to investigate in
a unified manner the corresponding Ward identities.  Since
the transformations generated by the $S_A$ may be non-linear,
they may get renormalized in the quantum theory.  To cope
with this feature, 
we extend the  approach of \cite{ZJ,BRS,Blasi} and introduce,
besides the standard antifields and local ghosts associated with
the gauge symmetry,
constant ghosts $\xi^A$ for {\em all} independent local conservation laws.
These constant ghosts are
assigned opposite ghost number and 
the same Grassmann parity as the corresponding generator
$S_A$ (consequently, the ghost number of $\xi^A$ equals the
order of the corresponding conservation law).  We then add the
term $S_A \xi^A$ to $S$ and search for a solution ${\cal S}[\Phi,
\Phi^*, \xi]$ of
the extended master equation
\begin{equation}
({\cal S},{\cal S}) + 2\sum_{r\geq 2}\frac 1{r!}\,
\frac{\partial^R {\cal S}}{\partial \xi^B} \, f_{A_1\cdots A_r}^{B}
\xi^{A_r}\cdots\xi^{A_1} = 0
\label{extendedmaster1}
\end{equation}
of the form
\begin{equation}
{\cal S} = S + S_A \xi^A + \sum_{r\geq 2}
\frac 1{r!}\, S_{A_1\cdots A_r}\xi^{A_r}\cdots\xi^{A_1},
\label{formofS}
\end{equation}
where the $f_{A_1\cdots A_r}^{B}$ are
the structure constants and the $S_{A_1\cdots A_r}$ are the local
functionals of the symmetry algebra described above.

The existence-proof of ${\cal S}$ becomes straightforward
if constant antifields $\xi_A^*$ conjugate to the 
constant ghosts $\xi^A$
are introduced through
\begin{equation} {\cal S}'=
{\cal S} + \sum_{r\geq 2}\frac 1{r!}\,\xi^*_B\, f_{A_1\cdots A_r}^{B}
\xi^{A_r}\cdots\xi^{A_1}.
\label{formofS'}
\end{equation}
With the additional antifields, the extended master equation 
(\ref{extendedmaster1}) takes
the familiar form  
\begin{equation}
({\cal S}', {\cal S}')' = 0,
\label{extendedmaster2}
\end{equation}
where the extended antibracket $(\ ,\ )'$ is given by
\begin{eqnarray*}
 (X,Y)'=
 \frac{\partial^R X}{\partial\xi^A}
 \frac{\partial^L Y}{\partial\xi^*_A}-
 \frac{\partial^R X}{\partial\xi^*_A}
 \frac{\partial^L Y}{\partial\xi^A}
 +\int \mbox{d}^nx \left[
 \frac{\delta^R X}{\delta\Phi^a(x)}
 \frac{\delta^L Y}{\delta\Phi^*_a(x)}-
 \frac{\delta^R X}{\delta\Phi^*_a(x)}
 \frac{\delta^L Y}{\delta\Phi^a(x)}\right].
\end{eqnarray*}
The construction of solutions to the familiar Eq.(\ref{extendedmaster2})
follows the standard pattern of homological
perturbation theory \cite{HPT2} (section 10.5.4), based  
on the crucial property that the Koszul-Tate differential
associated with both the local and the global symmetries is acyclic
in the appropriate functional space. Details will be given elsewhere.

A remarkable feature of the extended master equation (\ref{extendedmaster1})
or (\ref{extendedmaster2})
is that it encodes the structure constants of the algebra
of rigid symmetries described above.  Indeed, given $S$ and the
generators $S_A \xi^A$ (or actually, just their pieces linear
in the ghosts), the higher order functionals $S_{A_1 \dots A_r}$
and the structure constants $f^B_{A_1\dots A_r}$ are recursively
determined by the demand that the extended master 
equation be satisfied. This parallels the property of the usual 
master equation that encodes all the information on
the algebra of gauge transformations, including the Jacobi identities
of first and higher order \cite{bv,HPT2}.
When there are only standard rigid symmetries, Eq.(\ref{extendedmaster1}) 
reduces to the extended master
equation of \cite{Blasi},
\begin{equation}
({\cal S},{\cal S}) + 
\frac{\partial^R {\cal S}}{\partial \xi^C} \, f_{BA}^{C}
\xi^{A}\xi^{B} = 0.
\label{extended'}
\end{equation}

{}From the extended master equation (\ref{extendedmaster1}),
the Ward identitites for the Green functions can be derived as follows.
The generating functional for the Green functions of the
theory is given by the path integral
\[
Z_{J,K,\xi} = \int [D\Phi] \exp i \{  {\cal S}^{\Psi}[\Phi, K, \xi] + \int
d^n x J_a(x) \Phi^a(x) \}.
\]
The functional $ {\cal S}^{\Psi}[\Phi, K, \xi]$ appearing in $Z_{J,K,\xi}$ is
obtained from ${\cal S}[\Phi, \Phi^*, \xi]$
by making the transformation
$\Phi^*_a = K_a + \frac{\delta \Psi}{\delta \Phi^a}$, where the
gauge-fixing fermion $\Psi[\Phi]$ is chosen such that
${\cal S}^{\Psi}[\Phi, 0, 0]$ is completely gauge-fixed. 
The functional ${\cal S}^{\Psi}[\Phi, K, \xi]$ obeys
the same equation (\ref{extendedmaster1}) -- with $\Phi^*$ replaced by $K$ --
as ${\cal S}[\Phi,\Phi^*,\xi]$, because
the transformation from $\Phi^*$ to $K$ is a canonical transformation
that does not involve the $\xi^A$.  The fields
$J_a(x)$ and $K_a(x)$, as well as the constant ghosts $\xi^A$,
are external sources not to be integrated over in the path integral.
Now, perform in $Z_{J,K,\xi}$ the infinitesimal change of integration
variables
\begin{equation}
\Phi^a \rightarrow \Phi^a + ( \Phi^a,{\cal S}^{\Psi}) =
\Phi^a + \frac{\delta^L {\cal S}^{\Psi}}
{\delta K_a}.
\end{equation}
Using (\ref{extendedmaster1}) and assuming the measure
to be invariant \cite{comment}, the following Ward identity results
for $Z_{J,K,\xi}$:
\begin{equation}
\int d^n x\, J_a(x)\, \frac{\delta^L Z_{J,K,\xi}}{\delta K_a(x)}
- \sum_{r\geq 2}\frac 1{r!}\, 
\frac{\partial^R Z_{J,K,\xi}}{\partial \xi^B} \, f_{A_1\cdots A_r}^{B}
\xi^{A_r}\cdots\xi^{A_1}  = 0.
\label{WI1}
\end{equation}
Since Eq.(\ref{WI1}) is a linear functional
equation on $Z_{J,K,\xi}$, the generating functional
$W = -i \ln Z_{J,K,\xi}$ for the connected Green functions
obeys the same identity.
Performing the standard Legendre transformation
\begin{eqnarray}
& &\Gamma[\Phi_c, K, \xi] = W[J,K,\xi] - \int d^n x J_a(x) \Phi^a_c(x)\ ,
\nonumber \\
& &\Phi^a_c(x) = \frac{\delta^L W}{\delta J_a(x)}\, , \; \;
J_a(x) = - \frac{\delta^R \Gamma}{\delta \Phi^a_c(x)}
\label{Gamma}
\end{eqnarray}
one finds that the effective action $\Gamma$
fulfills
a Ward identity of the same form as (\ref{extendedmaster1}),
\begin{equation}
 \int \mbox{d}^nx\, 
 \frac{\delta^R \Gamma}{\delta\Phi^a_c(x)}\,
 \frac{\delta^L \Gamma}{\delta K_a(x)} +
 \sum_{r\geq 2}\frac 1{r!}\,
\frac{\partial^R \Gamma}{\partial \xi^B} \, f_{A_1\cdots A_r}^{B}
\xi^{A_r}\cdots\xi^{A_1} = 0.
\label{WI3}
\end{equation}

The  Ward  identities (\ref{WI1}, \ref{WI3})
capture the consequences of both the local
and the global symmetries for the generating functionals.
They hold even when the gauge fixing fermion is not invariant under 
the rigid symmetries, provided
there are no anomalies (see below).
The identities on the Green functions are obtained in the usual manner,
by differentiating
(\ref{WI1}, \ref{WI3})
with respect to the sources and setting these sources equal to zero
afterwards.

We have thus shown in this letter that the explicit inclusion of the
higher order rigid symmetries in the Ward identities avoids
the obstructions that may be encountered when trying to construct a solution
of the extended master equation (\ref{extended'}) that does not
take these higher order conservation laws into account.  While
(\ref{extendedmaster1}) is never obstructed \cite{comment3},
(\ref{extended'}) may
fail to have local solutions in the presence of higher
order symmetries.  Of course, (\ref{extended'})
will not be obstructed if a subset of first order conservation
laws is used that defines a subalgebra in the above sense. 
But otherwise  obstructions may arise
\cite{global}.  Furthermore,  the
Ward identities associated with the 
higher order conservation laws may yield
useful information on the Green functions.  
For instance, in the simple case of a $2$-form
abelian gauge field $B_{\mu \nu}$ with Lagrangian ${\cal L} =
{\cal L}(H_{\lambda \mu \nu}, \partial_\alpha H_{\lambda \mu \nu},
\dots)$, $H_{\lambda \mu \nu} = \partial_{[\lambda} B_{\mu \nu]}$, 
the Ward identity associated with the conservation law of
order three $\partial_\alpha (\delta {\cal L} / \delta H_{\alpha \mu \nu})
\approx 0$   can be brought under appropriate redefinitions to the form
\begin{equation}
\int d^nx\, \frac{\delta \Gamma }{\delta \eta(x)} = 0,
\label{example}
\end{equation}
where $\eta(x)$ is the ghost of ghost of order two 
associated with the local reducibility 
of the gauge symmetry, and where $\Gamma = \Gamma(K=0, \xi =0)$.  
The identity
(\ref{example}) expresses the invariance of the effective action
$\Gamma$ under constant
shifts of the ghost of ghost $\eta(x)$.

We close this letter by observing that the anomalies in the
rigid symmetries of all orders can be analysed along the algebraic lines
initiated in the pioneering work
\cite{BRS}.  Following the standard argument, an
anomaly appears as a
violation of the master equation (\ref{WI3}) for
the regularized $\Gamma$
and must fulfill, to lowest loop order,
the generalized Wess-Zumino consistency condition
\cite{wz}
\begin{equation}
D \int d^nx \, a = 0
\label{consistency}
\end{equation}
because $D^2=0$.  Here, the extended BRST differential $D$ is defined by
$D X = (X,{\cal S}')'$.  For local functionals not
involving the $\xi^*_A$, like $\int d^nx \, a$, $D$ takes
the form
\begin{equation}
DX = (X,{\cal S}) + \sum_{r\geq 2}\frac 1{r!}\,
\frac{\partial^R X}{\partial \xi^B} \, f_{A_1\cdots A_r}^{B}
\xi^{A_r}\cdots\xi^{A_1}.
\end{equation}
An anomaly (for the local or global symmetries, or combinations thereof)
is a solution of (\ref{consistency}) that is non-trivial, i.e. not of
the exact form $D\int  d^n x \, b$.  The investigation of the possible
anomalies in the local and global symmetries of all orders is
accordingly equivalent to the problem of computing the
cohomology of $D$ at ghost number one.

{\em Acknowledgements:}
This work has been supported  in part by research
funds from the F.N.R.S.
(Belgium) and research contracts with the Commission of the European
Community. F.B. has been supported by the Spanish
ministry of education (MEC).


\begin{thebibliography}{60}
\bibitem{comment0}
A conserved antisymmetric tensor is trivial if it is of the form
$j^{\mu_1 \dots \mu_k}
\approx \partial_\nu k^{\nu \mu_1 \dots \mu_k}$
where $k^{\nu \mu_1 \dots \mu_k}$ is also completely antisymmetric.
\bibitem{comment} If the measure is not invariant, one must replace
the extended master equation (\ref{extendedmaster1}) by the
quantum extended master equation, which reads
$$({\cal S},{\cal S}) + 2 i \Delta {\cal S}+ 2\sum_{r\geq 2}\frac 1{r!}\,
\frac{\partial^R {\cal S}}{\partial \xi^B} \, f_{A_1\cdots A_r}^{B}
\xi^{A_r}\cdots\xi^{A_1} = 0$$
where of course a meaningful (regularized) definition has to be
given to $\Delta$.  The Ward identitites
(\ref{WI1}) and (\ref{WI3}) are unchanged in absence
of anomalies.
\bibitem{comment3} Note that it is crucial, in order to avoid the
obstructions, to take the higher order conservation laws into account as we
have done here, with constant ghosts of ghost number $k$
(it would not work to 
treat the higher order conservations laws (\ref{cons2})
as ordinary conservation laws parametrized by further
indices and to associate with them constant ghosts
$\xi^{\mu_1 \dots\mu_k}$ of ghost number one).  Note also that
the scheme given in the example of \cite{global} for removing the obstruction
is just a particular case of the general method given here.
\end{thebibliography}
\end{document}